\documentclass[12pt, a4paper]{article}
\pdfoutput=1

\usepackage[a4paper,left=1.5cm,right=1.5cm,top=3cm,bottom=3cm]{geometry}
\usepackage{bmpsize}
\usepackage{amssymb,amsmath,amsfonts}
\usepackage[dvipsnames]{xcolor}
\usepackage{graphicx}
\usepackage{longtable}
\usepackage{verbatim}
\usepackage{color}

\usepackage{color}
\usepackage{hyperref}
\hypersetup{colorlinks, citecolor=bluscuro, linkcolor=black, urlcolor=bluscuro}
\definecolor{rossos}{cmyk}{0,1,1,0.55}
\definecolor{bluscuro}{rgb}{0.15, 0.2, .85}
\definecolor{bluchiaro}{cmyk}{1,.3,0.,0.1}

\graphicspath{{./Figures/}}
\def\bma#1{\mbox{\boldmath{$#1$}}}

\newcommand{\be}{\begin{equation}}
\newcommand{\ee}{\end{equation}}
\newcommand{\bea}{\begin{eqnarray}}
\newcommand{\eea}{\end{eqnarray}}
\newcommand{\beq}{\begin{equation}}
\newcommand{\eeq}{\end{equation}}
\newcommand{\dd}{{\rm d}}
\def\beqa{\begin{eqnarray}}

\def\eeqa{\end{eqnarray}}

\def\lsim{\mathrel{\rlap{\lower4pt\hbox{\hskip0.5pt$\sim$}}
    \raise1pt\hbox{$<$}}}         
\def\gsim{\mathrel{\rlap{\lower4pt\hbox{\hskip0.5pt$\sim$}}
    \raise1pt\hbox{$>$}}}         

\usepackage[normalem]{ulem}

\begin{document}

\begin{titlepage}
\begin{flushright}
\end{flushright}
\vspace{1cm}
\begin{center}
{\Large\bf\color{black} The Tunneling Potential Approach to Q-Balls}\\
\bigskip\color{black}
\vspace{1cm}{
{\large J.~R.~Espinosa$^1$}, {\large J.~Heeck$^2$} and {\large M. Sokhashvili$^2$}
\vspace{0.3cm}
} 
\\[7mm]
{\it $^1$ Instituto de F\'{\i}sica Te\'orica UAM/CSIC, \\  C/ Nicol\'as Cabrera 13-15, Campus de Cantoblanco, 28049, Madrid, Spain}\\[3mm]
{\it $^2$ Department of Physics, University of Virginia,
Charlottesville, Virginia 22904-4714, USA}
\end{center}
\bigskip

\vspace{.4cm}

\begin{abstract}

$Q$-balls are bound-state configurations of complex scalars stabilized by a conserved Noether charge $Q$. They are solutions to a second-order differential equation that is structurally identical to Euclidean vacuum-decay bounce solutions in three dimensions. This enables us to translate the recent \emph{tunneling potential approach} to $Q$-balls, which amounts to a reformulation of the problem that can simplify the task of finding approximate and even exact $Q$-ball solutions.

\end{abstract}
\bigskip

\end{titlepage}


\section{ Introduction \label{sec:Intro}} 

Classical field theory of massive bosons can admit stable, localized solutions, so called non-topological solitons~\cite{Lee:1991ax}. This requires an attractive force to bind the particles together and a conserved Noether charge $Q$ to forbid decays and self-annihilation. The minimal example is a complex scalar field $\Phi(\vec{x},t)$ with a $U(1)$-invariant potential $U(|\Phi|)$ that grows slower than $|\Phi|^2$ in some range. In that case, spherically symmetric localized solutions to the Klein--Gordon equation of the form $\Phi = \phi(|\vec{x}|) e^{i\omega t}/\sqrt2$ exist that can have the lowest energy per charge~\cite{Coleman:1985ki}.
We shall denote all such objects as $Q$-balls in a slight generalization of Coleman's definition.
Similar solitons can arise in multi-field scenarios and can be tackled with a similar approach as pursued below~\cite{Espinosa:2018szu}.

$Q$-balls are interesting examples of bound states that have been discussed in many different contexts. While the Standard Model of particle physics does not provide any elementary complex scalars that could condense into $Q$-balls, many extensions do. For example, supersymmetry naturally provides complex scalars (sfermions) charged under global $U(1)$ symmetries (baryon and lepton number) with attractive interactions in their potential, thus allowing for $Q$-balls~\cite{Kusenko:1997zq}. These stable objects could form and grow in the early universe and make up dark matter~\cite{Kusenko:1997si}.

Mathematically, $Q$-balls are described by a second-order differential equation for their radial profile $\phi(|\vec{x}|)$~\cite{Coleman:1985ki}. Equation and boundary conditions are structurally identical to Euclidean vacuum-decay bounce solutions in $d=3$ dimensions~\cite{Coleman:1977py}, as originally noted by Coleman.
We can therefore adapt the recent reinterpretation and approximation of bounces by one of the authors (JRE)~\cite{Espinosa:2018hue} to the $Q$-ball case, yielding a novel method to find exact and approximate soliton solutions.

The so-called tunneling potential approach \cite{Espinosa:2018hue} to the calculation of tunneling actions (which control the decay of false vacua) is an alternative to the usual Euclidean approach. The new method formulates the problem as a minimization problem for an action functional  $S[V_t]$ (defined in field space) that depends on the so-called tunneling function, $V_t(\phi)$, which connects the false vacuum and the basin of the true vacuum and describes the decay. The monotonicity of $V_t$ and the fact that $S[V_t]$ is a minimum (rather than a saddle point) make the new approach powerful for numerical applications, both for single-field \cite{Espinosa:2018hue} and multi-field \cite{Espinosa:2018szu} potentials in arbitrary dimensions. Moreover, the method leads in a natural way to a new procedure to find potentials which allow for exact solutions of the decay process. These appealing properties of the tunneling potential formalism can be applied to the study of $Q$-balls and is the main goal of this paper.

We introduce $Q$-ball terminology and notation in Sec.~\ref{sec:Qballs} and the reformulation in the tunneling approach in Sec.~\ref{sec:Vt}.
Approximations for the new sought-after function $\mathcal{E}(\phi)$ are derived in Sec.~\ref{sec:Eestimate} and then applied and compared to numerical solutions in Sec.~\ref{sec:comparison}.
In Sec.~\ref{sec:Exact} we use the new formalism to construct potentials that lead to exactly solvable $Q$-ball solutions.
We conclude in Sec.~\ref{sec:Conclusion}.
App.~\ref{app:flat} is devoted to a discussion of the large $Q$-ball limit in flat potentials, appropriate for potentials that do not have a thin-wall limit.

\section{ \texorpdfstring{$\bma{Q}$}{Q}-Ball Basics \label{sec:Qballs}} 

\begin{figure}[t!]
\begin{center}
\includegraphics[width=0.47\textwidth]{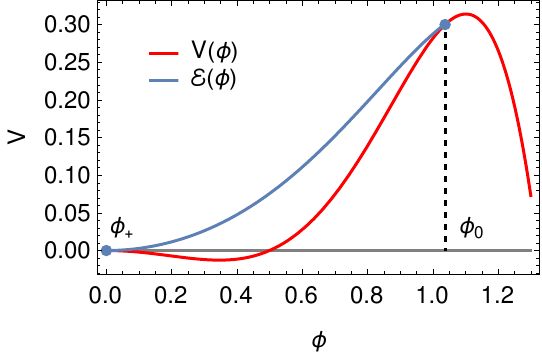}\,
\includegraphics[width=0.46\textwidth]{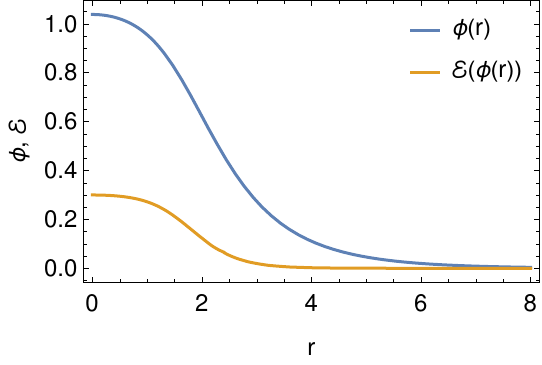}
\end{center}
\caption{Left: example of a $Q$-ball potential $V(\phi)$ (red), in which a particle rolls from $\phi_0$ to $\phi_+=0$ under friction. $\mathcal{E} (\phi)$ (blue) corresponds to the particle's total energy.
Right: the corresponding solution $\phi(r)$ as well as the change of total energy $\mathcal{E} $ with ``time'' $r$. 
\label{fig:potential}
}
\end{figure}

Using the notation established in the introduction, the $O(3)$-symmetric radial $Q$-ball profile $\phi(r)$ satisfies a differential equation of the form
\be
\ddot{\phi} +\frac{2}{r}\dot{\phi} =- V'\ ,
\label{EoMphi}
\ee
where a dot (prime) represents a derivative with respect to $r$ ($\phi$), 
with the boundary conditions
\be
\dot\phi(0)=0\ ,\quad \phi(\infty)=\phi_+\ .
\ee
Here, $\phi_+$ is a local maximum of $V$,  where the $U(1)$ symmetry is unbroken, which we could set to zero without loss of generality and $V(\phi)$ is an effective potential relevant for the description of $Q$-balls. This $V$ is different from the fundamental potential for the scalar field appearing in the Lagrangian, $U(|\Phi|)$, with 
\be
V(\phi)\equiv\frac12 \omega^2\phi^2 - U\left(\frac{\phi}{\sqrt2}\right) ,
\ee
where $\omega$ is the rotation frequency in internal space of the $Q$-ball solution.
An example for a possible potential is shown in Fig.~\ref{fig:potential} (left).

If we identify $r$ with time, Eq.~(\ref{EoMphi}) describes the motion of a particle with position $\phi$ in the potential $V(\phi)$, with a velocity and time-dependent friction force, left to roll from some initial position $\phi(0)=\phi_0$. The solution can be found by the undershoot/overshoot method, changing $\phi_0$ until the boundary condition at $r\rightarrow \infty$ is satisfied~\cite{Coleman:1985ki}.\footnote{Notice that this $V$ has the opposite sign of a vacuum-decay potential~\cite{Coleman:1977py}.}
We shall focus on ground-state solutions, for which $\phi (r)$ decreases monotonically.
The example of Fig.~\ref{fig:potential} illustrates the main qualitative features of $Q$-ball potentials and profiles.

The Noether charge $Q$ counts the number of $\phi$ particles minus antiparticles and for a $Q$-ball is given by the integral
\be
Q[\phi]=4\pi\omega\int_0^\infty  \phi^2 r^2dr\ .
\ee
For a fixed charge $Q$, the $Q$-ball profile $\phi(r)$ minimizes the energy functional 
\be
E[\phi] =\omega Q + 4\pi\int_0^\infty  \left[\frac12 \dot{\phi}^2 - V(\phi)+V(\phi_+)\right]r^2dr\ .
\label{EQ}
\ee
If $\phi$ satisfies the equation of motion, energy and charge are related through the differential relationship
\be
\frac{d E}{d\omega} =\omega \frac{d Q}{d\omega } \,,
\label{dEdQ}
\ee
which also allows for an interpretation of $\omega = d E/d Q$ as the chemical potential~\cite{Friedberg:1976me,Lee:1991ax}. We require $0\leq \omega_0 < \omega < m_\phi$ for localized solutions, with $\omega \sim \omega_0$ corresponding to the limit of large $Q$-balls, defined below.

We can distinguish two qualitatively different potentials and resulting solitons: 
\begin{enumerate}
\item[(A)] $V(\phi)$ has a global maximum at some $\phi_- > \phi_+$, as the one illustrated in Fig.~\ref{fig:potential}.  The particle can then spend a long time near this point before starting to roll, which implies an essentially constant $\phi(r)\simeq \phi_-$ up to some potentially large radius $R_c$~\cite{Coleman:1985ki}. Solitons with such a thin-wall limit were dubbed $Q$-balls by Coleman.
Here, $\omega_0$ is the value of $\sqrt{U(|\Phi|)/|\Phi|^2}$ at its minimum, which occurs at the non-zero finite value $|\Phi| \simeq \phi_-/\sqrt2$~\cite{Coleman:1985ki}.

\item[(B)] $V(\phi)$ does not have a global maximum but rather grows with $\omega^2\phi^2/2$ for large $\phi$. This occurs for $U$ potentials that are approximately flat and predate Coleman's $Q$-balls~\cite{Rosen:1968mfz,Dvali:1997qv}. Here, the particle starts to roll immediately and the only way to make the $Q$-ball larger is to start $\phi$ at higher and higher values.
In the vacuum-decay analogy, this corresponds to tunneling into an unbounded vacuum.
Here, $\omega_0$ is simply zero.
\end{enumerate}
Notice that neither case can be realized with a renormalizable bounded-from-below single-field potential $U(|\Phi|)$, but both can be excellent \emph{effective} descriptions of scenarios arising in renormalizable multi-field cases.
Both cases allow for arbitrarily large stable $Q$-balls that can be approximated with relative ease by using the large radius as an expansion parameter.

In the thin-wall limit, the potential difference between the maxima, $\Delta V\equiv V(\phi_-)-V(\phi_+)$, is small compared to the depth of the minimum between them 
and the profile has a sharp transition between $\phi_0\simeq \phi_-$ for
$r<R_c$ and $\phi_+$ for $r>R_c$, where the bubble radius $R_c$
can be calculated analytically in terms of the surface tension 
\be
\sigma\simeq \int_{\phi_+}^{\phi_-}\sqrt{2[V(\phi_-)-V(\phi)]}d\phi\ ,
\ee 
as $R_c=2\sigma/\Delta V$. In this thin-wall case the $Q$-ball charge and energy can be obtained analytically  as
\be
Q_{tw}=\frac{4\pi}{3}\omega\phi_-^2R_c^3\ ,\quad
E_{tw}=\omega Q+\frac{16\pi}{3}\frac{\sigma^3}{\Delta V^2}\ ,
\label{EQtw}
\ee
where $\omega\sim \omega_0$ is assumed to be non-zero and $\sigma^3/\Delta V^2\propto R_c^2$ corresponds to the surface energy. For large $Q$-balls, stability is ensured because $E \simeq \omega_0 Q < m_\phi Q$~\cite{Coleman:1985ki}.
The thin-wall limit works remarkably well for $Q$-balls of type A and can be systematically improved order by order in $1/R$~\cite{Heeck:2020bau,Heeck:2022iky}.

Large $Q$-balls of type B in flat potentials are even simpler to describe because Eq.~\eqref{EoMphi} becomes a linear differential equation for large $\phi$ with solution $\sin (\omega r)/(\omega r)$. This leads to a markedly different scaling in the large-$R$ limit:
$Q \propto R^{4} \propto E^{4/3}$~\cite{Dvali:1997qv}.
Once again large $Q$-balls are stable because $E/(m_\phi Q) \propto Q^{-1/4}  $ becomes smaller than one beyond some critical charge.
We provide the large $Q$-ball solution for this case in App.~\ref{app:flat} for the convenience of the reader since it is not as widely known as the thin-wall limit.

\section{\texorpdfstring{$\bma{Q}$}{Q}-Balls via the Tunneling Potential Approach \label{sec:Vt}} 

The application of the so-called tunneling potential approach to the study of $Q$-balls is based on the fact that the equation for the $Q$-ball profile, (\ref{EoMphi}), has the form of a Coleman bounce equation in three dimensions. We can then borrow directly known $V_t$ results and properties~\cite{Espinosa:2018hue}.
Notice that, in what follows, there is a flip in the signs of both $V$ and $V_t$ with respect to the tunneling-potential literature in order to match the standard $Q$-ball notation.

With these sign flips in mind, the key quantity in the tunneling potential approach is 
\be
V_t(\phi)\equiv \frac12 \dot\phi^2 + V(\phi) \ ,
\label{Vt}
\ee
where it is understood that $\dot\phi$ above is expressed in terms of the field $\phi$ via the $Q$-ball profile.
$V_t$ is nothing but the total particle energy in the mechanics analogy, which is monotonically decreasing due to the friction term in the equation of motion.
From now on, we use $\mathcal{E}(\phi)$ instead of $V_t$ to make  this interpretation explicit.

The entire $Q$-ball problem can be rewritten in terms of $\mathcal{E}(\phi)$ rather than $\phi(r)$.
$\mathcal{E}(\phi)$ connects the local maximum at $\phi_+$ to some $\phi_0$ such that the $Q$-ball energy functional
\be
\boxed{E[\mathcal{E}]=\omega Q+\frac{16\pi}{3} \int_{\phi_+}^{\phi_0}\frac{[2(\mathcal{E}-V)]^{3/2}}{\left(\mathcal{E}'\right)^2}d\phi}
\label{EQVt}
\ee
is minimized (for fixed $Q$), reproducing the $Q$-ball energy in (\ref{EQ}). 
The integral here is the surface energy of the $Q$-ball and corresponds to the Euclidean action integral in the vacuum-decay analogy.
The total charge of the $Q$-ball can be written as 
\be
Q[\mathcal{E}]=16\pi\omega \int_{\phi_+}^{\phi_0}\frac{\sqrt{2(\mathcal{E}-V)}}{\left(\mathcal{E}'\right)^2}\phi^2d\phi\ .
\label{Qintegral}
\ee

One can remove altogether the reference to the $Q$-ball profile and real space in favor of $\mathcal{E}(\phi)$ as follows. From (\ref{Vt}),
\be
\dot\phi = - \sqrt{2[\mathcal{E}(\phi) - V(\phi)]}\ ,
\label{dphi}
\ee
where the minus sign, chosen due to $\phi_+<\phi_-$, should be a plus if $\phi_+>\phi_-$. Eq.~(\ref{dphi}) allows to remove  derivatives of $\phi$ in terms of $\mathcal{E}$ (and $V$). The radial coordinate $r$ can then be extracted from (\ref{EoMphi})
as
\be
r=2\sqrt{2(\mathcal{E} - V)/(\mathcal{E}')^2}\ .
\label{r}
\ee
Taking a derivative of the above with respect to $r$ we get a second-order differential equation for $\mathcal{E}$:
\be
\boxed{\left(3\mathcal{E}'-2 V' \right)\mathcal{E}' = 4(\mathcal{E}-V)\mathcal{E}''}\ ,
\label{VtEoM}
\ee
which also follows as the Euler--Lagrange equation from extremizing 
(\ref{EQVt}).\footnote{It can be proven~\cite{Espinosa:2018hue} that the ${\mathcal E}$ that solves this Euler--Lagrange equation {\em minimizes} the integral in (\ref{EQVt}) for monotonically increasing  ${\mathcal E}$'s.\label{Emin}} In the $\mathcal{E}$-formulation of the problem, one should find a $\phi_0$ and a $\mathcal{E}(\phi)$ that solve (\ref{VtEoM}) with boundary conditions
\be
\mathcal{E}(\phi_+)=V(\phi_+)\ ,\quad \mathcal{E}(\phi_0)=V(\phi_0)\ .
\label{BCV}
\ee
Assuming $V'(\phi_+)=0$, Eq.~(\ref{VtEoM}) also leads to
\be
\mathcal{E}'(\phi_+)=0\ ,\quad \mathcal{E}'(\phi_0)=2 V'(\phi_0)/3\ .
\label{BCVp}
\ee
At this point it is far from obvious why reformulating $Q$-balls in terms of $\mathcal{E}(\phi)$ is beneficial,  but it turns out to be quite simple to estimate $\mathcal{E}(\phi)$ for a given potential to get an accurate approximation to the $Q$-ball energy using (\ref{EQVt}).

Equation (\ref{EQVt}) can be regarded as a generalization of the thin-wall approximation (in the sense that  
the expression for the action involves just an integral in field space rather than in real space 
involving the $Q$-ball profile) that is valid for generic potentials.
In the limit of quasi-degenerate maxima we recover the
thin-wall result in a very direct way as follows: Given the properties of $\mathcal{E}$, for near degenerate maxima one has $\mathcal{E}'\ll |(V-\mathcal{E})'|$ -- except at small regions around the point $(V-\mathcal{E})'=0$ and at the maxima $\phi_\pm$, where $\mathcal{E}=V$ -- 
so (\ref{VtEoM})
can be reduced to the form $(\mathcal{E}-V)'\mathcal{E}'\simeq 2(\mathcal{E}-V)\mathcal{E}''$.
This can be readily integrated to get 
\be
\sqrt{2(\mathcal{E}-V)}\simeq C \mathcal{E}'\ ,
\label{tw}
\ee 
with an integration constant $C$. Integrating (\ref{tw}) in the interval $(\phi_+,\phi_0\simeq\phi_-)$ gives $C$ in terms of the wall tension $\sigma$ and the potential difference $\Delta V$ as
\be
\sigma\equiv
\int_{\phi_+}^{\phi_-}\sqrt{2[\mathcal{E}(\phi)-V(\phi)]}\, d\phi \simeq C \Delta V\ .
\label{sigmatw}
\ee
Plugging (\ref{tw}) and (\ref{sigmatw}) in the energy integral (\ref{EQVt}) reproduces the correct thin-wall energy (\ref{EQtw}).
Moreover, to get the critical bubble radius we just plug (\ref{tw}) in (\ref{r}) to get $r=R_c=2C$, and using (\ref{sigmatw}) we get the correct result, $R_c=2\sigma/\Delta V$.

To sum up, $\mathcal{E}(\phi)$ captures all the relevant information of the $Q$-ball profile and has the appealing property of being quite featureless [{\it e.g.} a potential with a bumpy barrier gives a $\dot\phi^2/2$ that also has bumps; both sets of bumps cancel out in $\mathcal{E}(\phi)$], as can be seen in Fig.~\ref{fig:potential}. This property can be exploited to estimate the $Q$-ball energy  as is shown next.

\section{Estimates of the  Energy\label{sec:Eestimate}} 
Let us give approximations for $\mathcal{E}$ of increasing level of sophistication. The simplest is the linear one:
\be
\mathcal{E}_{1}(\phi)=V_0\frac{\phi}{\phi_0}\ ,
\label{Vi1}
\ee
where $V_0=V(\phi_0)$, and it satisfies the boundary conditions (\ref{BCV}). To also satisfy the boundary condition (\ref{BCVp}) on $\mathcal{E}'(\phi_0)$, one can use the quadratic approximation
\be
\mathcal{E}_{2}(\phi)=\mathcal{E}_{1}(\phi)+\frac{\phi}{3\phi_0^2}\left(2\phi_0V'_0-3V_0\right)(\phi-\phi_0)\ ,
\label{Vi2}
\ee
where $V'_0=V'(\phi_0)$. One can go further, matching also $\mathcal{E}'(\phi_+)$ using the cubic approximation
\be
\mathcal{E}_{3}(\phi)=\mathcal{E}_{2}(\phi)+\frac{2\phi}{3\phi_0^3}\left(\phi_0V'_0-3V_0\right)(\phi-\phi_0)^2\ .
\label{Vi3}
\ee
So far, the previous approximations focus only on the boundary conditions at $\phi_+$ and $\phi_0$. We can add information about 
the local minimum between the maxima, forcing the  $\mathcal{E}$ approximation to satisfy Eq.~(\ref{VtEoM}) at $\phi_T$, the value at which $V(\phi)$ is minimal. This can be achieved with the quartic approximation 
\be
\mathcal{E}_{4}(\phi)=\mathcal{E}_{3}(\phi)+a_4 \phi^2(\phi-\phi_0)^2\ ,\label{Vi4}
\ee
where
\be
a_4=\frac{
a_{0T}+\sqrt{a_{0T}^2-
\phi_T^2\phi_0^2\phi_{0T}^2 U_{3T}}}{2\phi_T^2\phi_0^2\phi_{0T}^2}\ ,
\ee
with $\phi_{0T}\equiv \phi_0-\phi_T$,
$U_{3T}\equiv 3(\mathcal{E}'_{3T})^2+4(V_T-\mathcal{E}_{3T})\mathcal{E}''_{3T}$,
 $\mathcal{E}_{3T}\equiv \mathcal{E}_{3}(\phi_T)$ and
\bea
a_{0T}&=&
2(V_T-\mathcal{E}_{3T})(6\phi_T\phi_{0T}-\phi_0^2)-3\phi_T(\phi_{0T}-\phi_T)\phi_{0T}\mathcal{E}'_{3T}+\phi_T^2\phi_{0T}^2\mathcal{E}''_{3T} .
\eea
This approximation also satisfies the boundary conditions in Eqs. (\ref{BCV}) and (\ref{BCVp}). 

More generally, one could pick $a_4$ so that the differential equation is satisfied at some other value, say at $\phi_0/\eta$ with $\eta>1$ (or even leave $a_4$ as a free parameter and minimize $E$ with respect to $\phi_0$ and $a_4$). This might be necessary to make this approximation useful for type-B $Q$-balls, where $\phi_0$ can be much larger than $\phi_T$, so $\phi_T$ is not a useful reference point.
Matching at $\phi_\eta\equiv\phi_0/\eta$ gives
\be
a_4=\frac{
a_{0\eta}-\sqrt{a_{0\eta}^2-
\phi_\eta^2\phi_0^2\phi_{0\eta}^2 U_{3\eta}}}{2\phi_\eta^2\phi_0^2\phi_{0\eta}^2}\ ,
\label{eq:a4_eta}
\ee
with $\phi_{0\eta}\equiv \phi_0-\phi_\eta$,
$U_{3\eta}\equiv 3(\mathcal{E}'_{3\eta})^2+4(V_\eta-\mathcal{E}_{3\eta})\mathcal{E}''_{3\eta}-2V'_\eta \mathcal{E}'_{3\eta}$,
 $\mathcal{E}_{3\eta}\equiv \mathcal{E}_{3}(\phi_\eta)$, $V_\eta\equiv V(\phi_\eta)$ and
\bea
a_{0\eta}&=&
2(V_\eta-\mathcal{E}_{3\eta})(6\phi_\eta\phi_{0\eta}-\phi_0^2)+\phi_\eta\phi_{0\eta}(\phi_{0\eta}-\phi_\eta)(V'_\eta-3\mathcal{E}'_{3T})+\phi_\eta^2\phi_{0\eta}^2\mathcal{E}''_{3\eta} \,.
\eea
The optimal choice of $\eta$ depends on the potential $V$ and is particularly important for large $Q$-balls, where some $\eta$ can lead to unphysical, e.g.~non-monotonic,  $\mathcal{E}$.
For small $Q$-balls, the exact value for $\eta$ is much less important and the $\mathcal{E}_{4}$ approximation works remarkably well.  This makes the above procedure nicely complementary to the well-known large $Q$-ball approximations.

 For an approximate ansatz $\mathcal{E}_{a}(\phi)$,  (\ref{EQVt}) gives an approximate $Q$-ball energy $E_a$ bigger than the true $E$,  see footnote~\ref{Emin}. $E_a$ is a function of the end point $\phi_0$ whose minimum gives the best estimate for $E$ for the used ansatz.

As we will show in the next section, $\mathcal{E}_{4}$ already provides excellent approximations for typical $Q$-ball potentials. The direct outputs are $\phi_0$ and the integral in Eq.~\eqref{EQVt}, essentially $E-\omega Q$. The actual profile $\phi (r)$ can be obtained via $r(\phi)$ from Eq.~\eqref{r}, which also provides the $Q$-ball radius, e.g.~conveniently defined through $\phi (R) = \tfrac23 \phi_0$~\cite{Heeck:2022iky}. Lastly, the charge $Q$ is in principle defined through $\mathcal{E}$ in Eq.~\eqref{Qintegral}. However, for large $Q$-balls this integral is often not well approximated by this method.
Instead, one can make use of the $Q$-ball relation $d E/d \omega = \omega d Q/d \omega$ to write
\begin{align}
Q = \frac{d}{d\omega} \left( \omega Q - E\right) .
\label{eq:Qtrick}
\end{align}
This allows us to obtain the $Q$ integral directly from the well-approximated result for the surface energy $E-\omega Q$, which works well even for large $Q$-balls.
Unless noted otherwise, this is how we obtain $Q$ in the following.

\section{Comparison with Numerical Solutions}
\label{sec:comparison}

In the following we will compare the $\mathcal{E}$ approximations to numerical solutions of the $Q$-ball equations.

\subsection{Polynomial Potentials}

A large class of $Q$-balls of type A can be described by polynomial potentials,
\begin{align}
U(|\phi|) = m^2_\phi |\phi|^2 -\beta |\phi|^p+\xi |\phi|^q \,,
\label{eq:phi_potential}
\end{align}
with  $2 < p < q$ and positive coefficients $ m^2_\phi$, $\beta$, and $\xi$. The thin-wall limit of these cases has been discussed extensively in Refs.~\cite{Lennon:2021uqu,Heeck:2022iky}. By redefining parameters one can bring Eq.~\eqref{EoMphi} to the form
\begin{align}
f''(\rho) +\frac{2}{\rho} f'(\rho)  = - \frac{d}{d f(\rho)}\left[ \frac{(p-q) (\kappa^2-1) f(\rho)^2 - (q-2) f(\rho)^p + (p-2) f(\rho)^q}{2(p-q)}\right] ,
\label{eq:eomf}
\end{align}
with dimensionless radial coordinate $\rho$, dimensionless $Q$-ball profile $f(\rho)$, and $0<\kappa < 1$ playing the role of $\omega$, with $\kappa\to 0$ corresponding to the thin-wall limit.

\begin{figure}[t!]
\begin{center}
\includegraphics[height=0.34\textwidth]{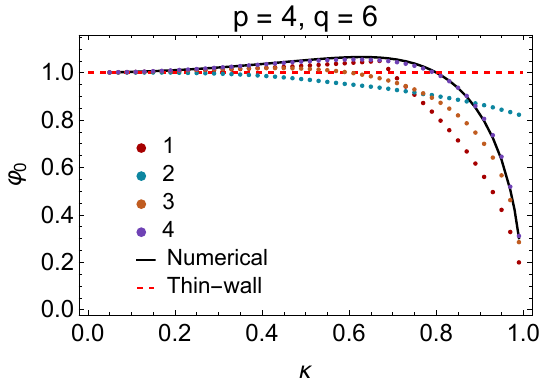}\,
\includegraphics[height=0.34\textwidth]{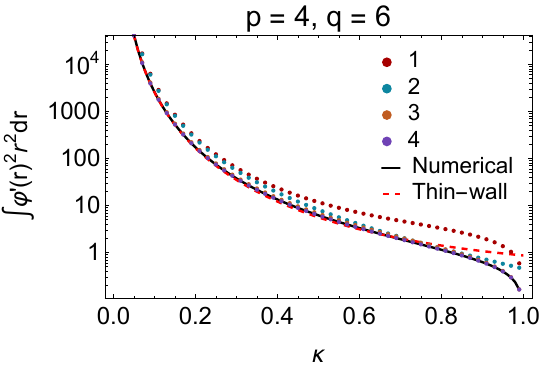}
\end{center}
\caption{Left: prediction of $\phi_0$ using the $\mathcal{E}_{a}$ approximations of Sec.~\ref{sec:Eestimate} for the potential from Eq.~\eqref{eq:eomf} with $(p,q) = (4,6)$, together with the actual value in black.
Right: corresponding values for the  surface-energy integral.
The red dashed line corresponds to the thin-wall prediction from Eq.~\eqref{eq:thinwall46}. 
\label{fig:polynomial_example}
}
\end{figure}

In Fig.~\ref{fig:polynomial_example} (left) we show the $\phi_0$ prediction for $(p,q) = (4,6)$ as a function of $\kappa$ for the four increasingly sophisticated approximations of $\mathcal{E}$ from the previous section, together with the true value obtained from numerically solving Eq.~\eqref{eq:eomf}~\cite{Heeck:2022iky}. We can see that the $\mathcal{E}$ predictions become increasingly better, with $\mathcal{E}_{4}$ reaching percent level accuracy. The corresponding predicted minimum of the surface-energy integral is shown in Fig.~\ref{fig:polynomial_example} (right) together with the true result and the thin-wall prediction~\cite{Heeck:2020bau}
\begin{align}
\int [f'(\rho)]^2 \rho^2 d\rho \simeq \frac{1}{4\kappa^2} \,.
\label{eq:thinwall46}
\end{align}
For the surface-energy integral, the $\mathcal{E}_{4}$ prediction is at the per-mille level, as is the prediction for the $Q$ integral using Eq.~\eqref{eq:Qtrick}.
The $Q$-ball radius is obtained from Eq.~\eqref{r} and for $\mathcal{E}_{4}$ agrees at the percent level with the true values.
Notice that the $\mathcal{E}$ approximation is not restricted to the thin-wall region of parameter space but works well even for small $Q$-balls, thus nicely complementing existing analytical thin-wall results.

Using the results from Ref.~\cite{Heeck:2022iky} we have confirmed the above qualitative picture for many other values of $p$ and $q$ in Eq.~\eqref{eq:eomf}. We find a similar excellent agreement between the $\mathcal{E}_{4}$ predictions and numerical results in all cases and thus expect this approach to be successful for all $Q$-balls of type~A.
When $p$ and $q$ take on very large values, the minimum of $V$ moves closer to $\phi_0$ and the $a_4$ matching at the minimum is no longer optimal. Instead, it becomes beneficial to match $a_4$ at $\phi_0/2$ using Eq.~\eqref{eq:a4_eta}. The so-obtained $\mathcal{E}_4$ is then an excellent approximation for essentially arbitrary $p$ and $q$; for large $Q$-balls, a thin-wall prediction has been given in Ref.~\cite{Heeck:2022iky}. We stress that for large $p$ and $q$ or $\kappa$ it is numerically challenging to solve the actual differential equations, so approximations of the kind shown here are very useful to study those $Q$-balls.

\subsection{Flat Potential}

Having shown that the $\mathcal{E}$ approach works remarkably well for $Q$-balls that have a thin-wall limit, let us investigate solitons that are qualitatively different and do not have a global $V$ maximum. 
An example for a flat $U$ potential that generates such $Q$-balls was given long ago by Rosen~\cite{Rosen:1968mfz},
\begin{align}
U(\phi) = 
 m_\phi^2 \Lambda^2 \left(1-e^{-|\phi|^2/\Lambda^2}\right) ,
\end{align}
and corresponds to the smoothed-out version of the  simple piecewise quadratic potential~\cite{Copeland:2009as}
\begin{align}
U(\phi) = \begin{cases}
m_\phi^2 |\phi|^2 \,, & \text{ for } |\phi(r)|\leq \Lambda\,,\\
m_\phi^2 \Lambda^2 \,, & \text{ for } |\phi(r)|> \Lambda\,,
\end{cases}
\label{eq:piecewise_quadratic}
\end{align}
argued to arise in gauge-mediated MSSM potentials~\cite{MacKenzie:2001av}. Another simple renormalizable realization of flat potentials can be found in the Friedberg--Lee--Sirlin model~\cite{Friedberg:1976me,Heeck:2023idx}.
Upon rescaling, this scenario can be brought to a similar form as above, involving only dimensionless quantities:
\begin{align}
f''(\rho) +\frac{2}{\rho} f'(\rho)  = -\frac{1}{2} \frac{d}{d f(\rho)}\left[ \kappa^2 f(\rho)^2 -1 + e^{-f(\rho)^2}\right] .
\label{eq:flateom}
\end{align}
Even though these solitons do not have a thin-wall limit, they become large and stable for small $\kappa$ with $\phi_0 \simeq \pi/\kappa$ and $\int [f'(\rho)]^2 \rho^2 d\rho\simeq \pi^3/2\kappa^3$, as shown in App.~\ref{app:flat}.

\begin{figure}[t!]
\begin{center}
\includegraphics[height=0.33\textwidth]{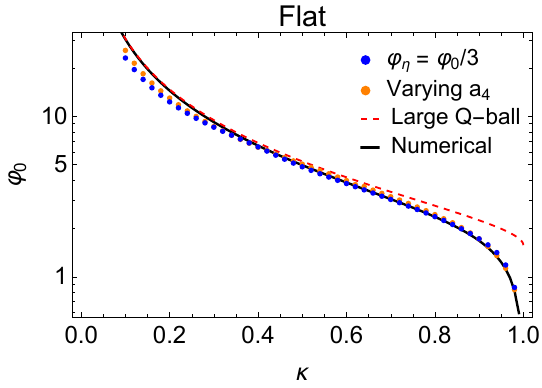}\,
\includegraphics[height=0.33\textwidth]{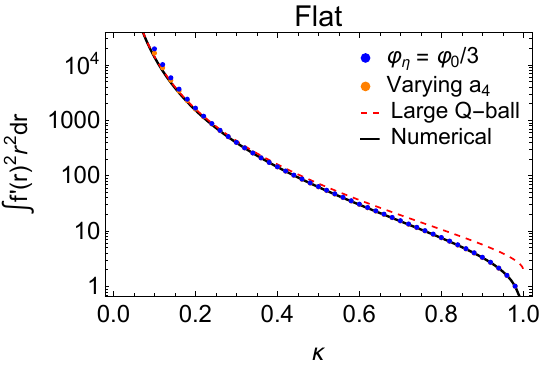}
\end{center}
\caption{Left: prediction of $\phi_0$ using the $\mathcal{E}_{4}$ approximations of Sec.~\ref{sec:Eestimate} for the potential from Eq.~\eqref{eq:flateom}, together with the actual value in black. Two different methods for $a_4$ are presented, one where $a_4$ matches the differential equation at $\phi_\eta = \phi_0/3$, and one where $a_4$ is varied freely to minimize the energy.
Right: corresponding values for the  surface-energy integral.
The red dashed line shows the large $Q$-ball approximation from App.~\ref{app:flat}.
\label{fig:flat_example}
}
\end{figure}

We restrict ourselves to the $\mathcal{E}_{4}$ prediction here. Since in this potential the position of the $V$ minimum is at $\phi_T = \sqrt{\log \kappa^{-2}}$, the distance to $\phi_0$ becomes arbitrarily large for small $\kappa$ and matching $a_4$ at the minimum is not ideal. Instead, we matched at $\phi_0/3$ and also left $a_4$ as a free minimization parameter, which is numerically more challenging but should give the best results.
In Fig.~\ref{fig:flat_example}, we compare these $\mathcal{E}_{4}$ predictions with the numerical results. As expected, leaving $a_4$ free gives the best results but matching at $\phi_0/3$ also gives a good approximation. For small $Q$-balls, both approaches for $\mathcal{E}_{4}$  work just as well as for thin-wall $Q$-balls, i.e.~with percent level accuracy. For small $\kappa$, i.e.~large $Q$-balls, the approximations become worse and should be replaced by the approximation derived in App.~\ref{app:flat}.
Alternatively, one could approximate $\mathcal{E}$ with higher-order polynomials to allow more freedom in the $\mathcal{E}$ shape.

\section{\texorpdfstring{Potentials with Exactly Solvable $Q$-Balls}{Potentials with Exactly Solvable Q-Balls} \label{sec:Exact}} 

Another application of the $\mathcal{E}$ approach is that it allows to generate in a straightforward manner potentials (or pieces of them) that lead to analytic solutions of the $Q$-ball problem. Instead of
starting from $V$ and solving for $\mathcal{E}$, one postulates a given 
$\mathcal{E}$ and  integrates (\ref{VtEoM}) to obtain the corresponding $V$ as
\be
V(\phi)=\mathcal{E}(\phi)+\frac12 [\mathcal{E}'(\phi)]^2
\int_{\phi_0}^\phi\frac{d\tilde\phi}{\mathcal{E}'(\tilde\phi)}\ .
\label{VfromVt}
\ee
When an approximate ansatz for $\mathcal{E}$ is used for a given potential $V(\phi)$, formula (\ref{VfromVt}) can also be used to check how close is the potential derived from the ansatz  to the original potential.

\subsection{Periodic \texorpdfstring{$\mathcal{E}$}{E}}

\begin{figure}[t!]
\begin{center}
\includegraphics[width=0.46\textwidth]{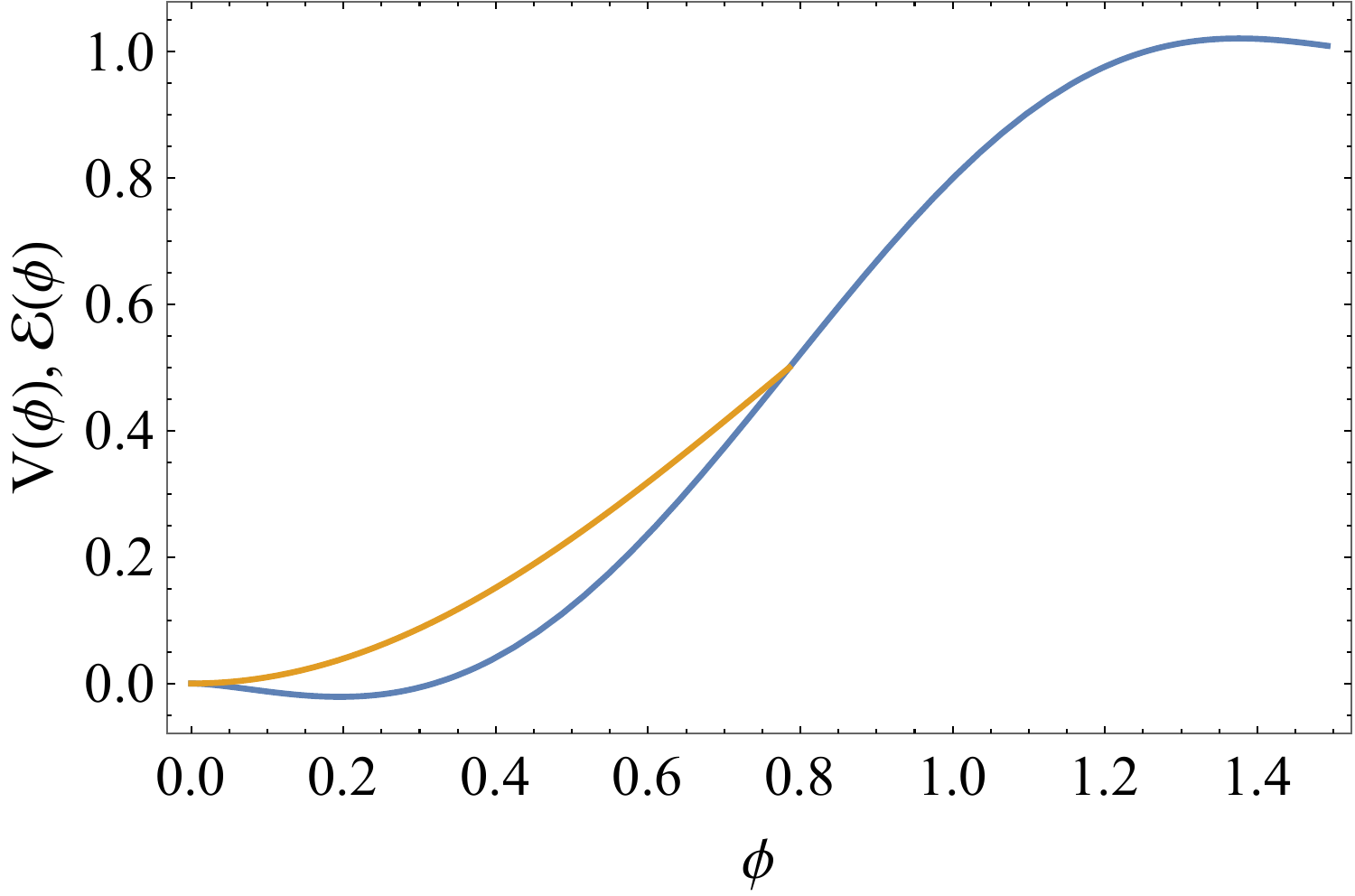}
\includegraphics[width=0.45\textwidth]{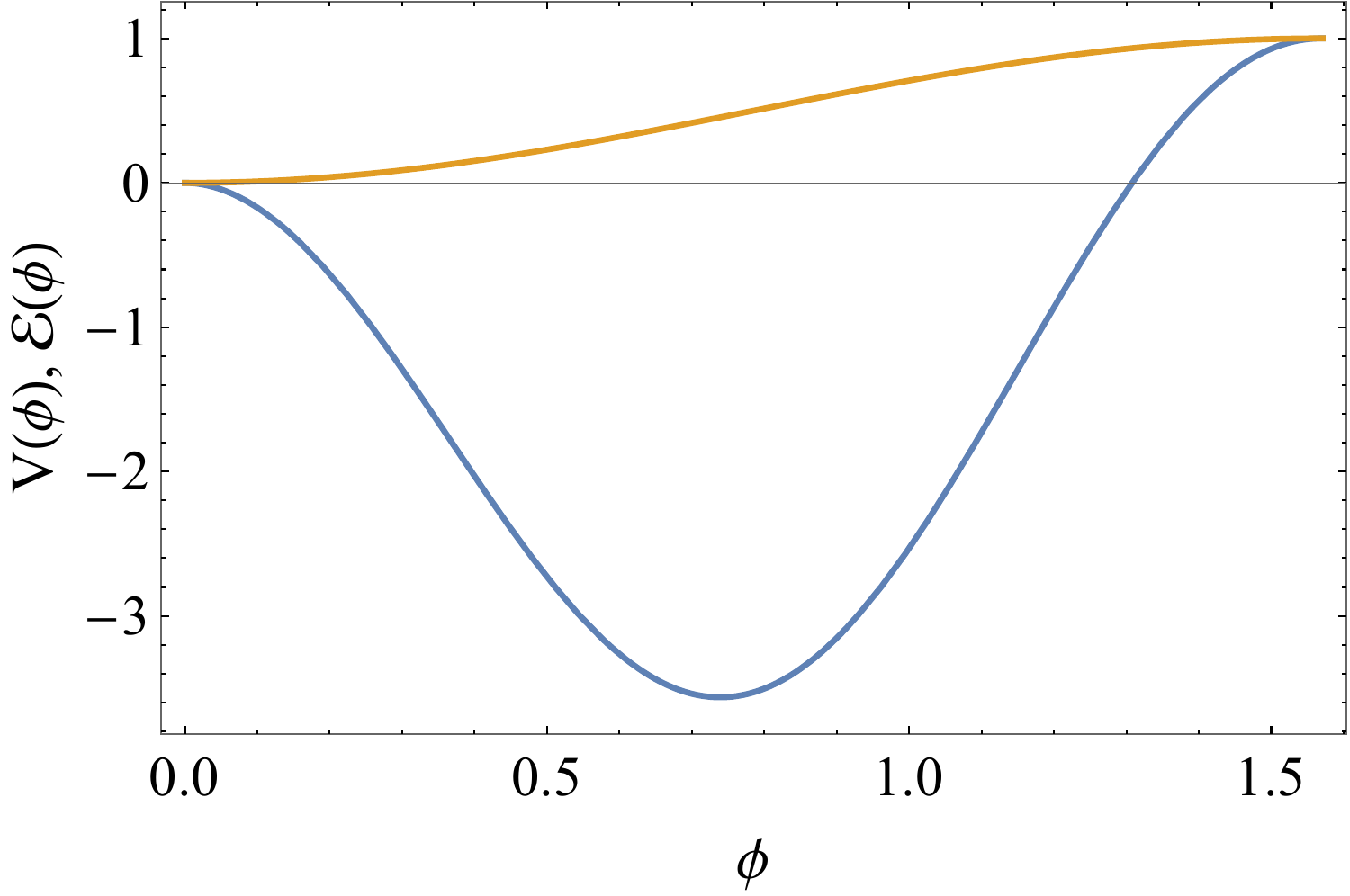}\\
\includegraphics[width=0.45\textwidth]{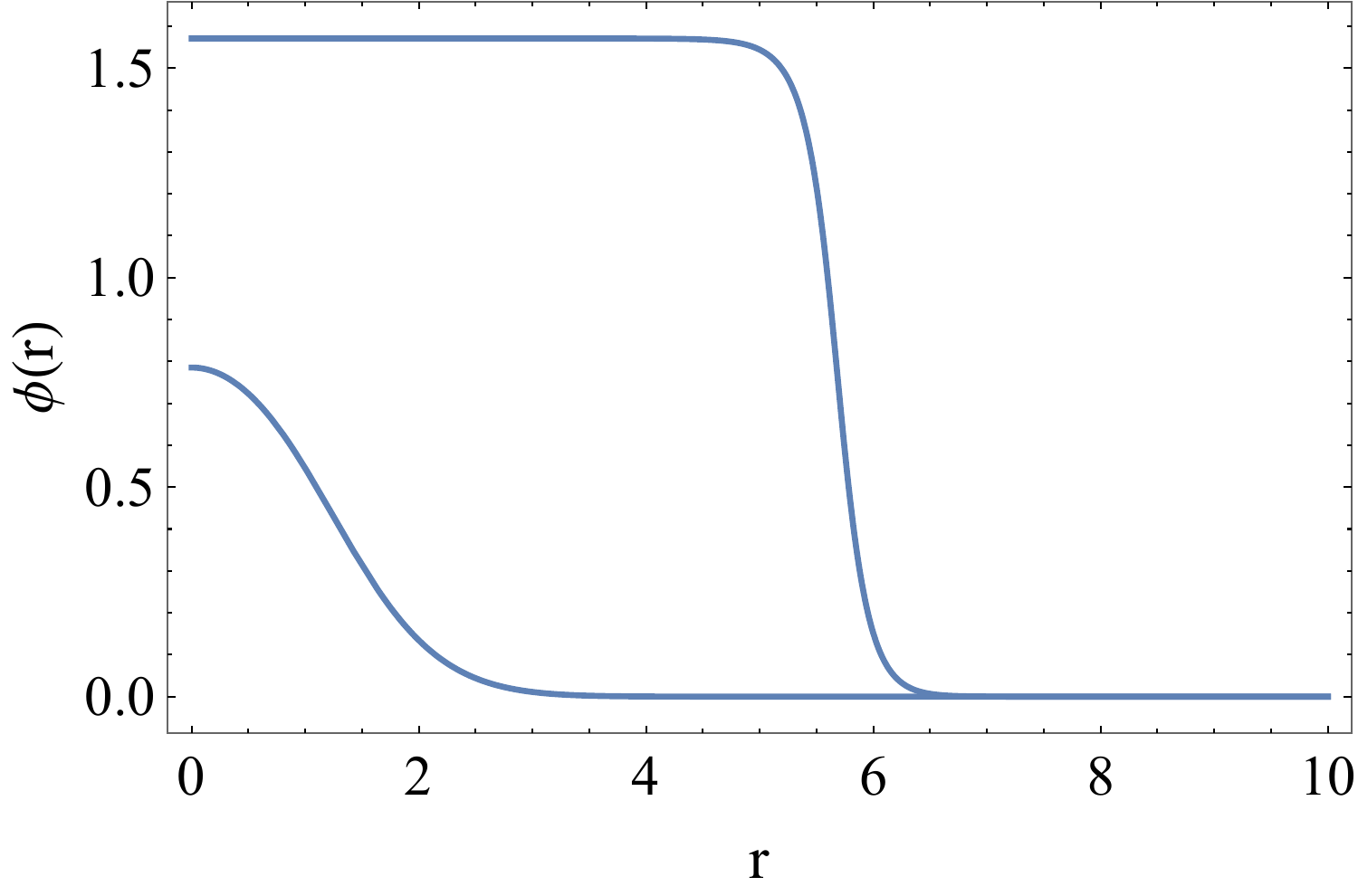}
\end{center}
\caption{For the analytic example of (\ref{Vtexample}), upper plots:  potential, $V$, (blue curve) and tunneling potential, $\mathcal{E}$ (orange curve) for $\phi_0=\pi/4$ (left plot)
and $\phi_0=\pi/2-10^{-7}$ (right plot). Lower plot:  corresponding $Q$-ball profiles $\phi(r)$.
\label{fig:VtV}
}
\end{figure}

As a first example, consider
\be
\mathcal{E}(\phi) = \sin^2\phi\ ,
\label{Vtexample}
\ee 
for simplicity in conveniently chosen units, although mass parameters can be put back easily if necessary.
Using (\ref{VfromVt}), the corresponding potential is
\be
V(\phi) =\sin^2\phi\ \left[1+\cos^2\phi\ \log\frac{\tan\phi}{\tan\phi_0}\right]
 .
\ee
The profile of the $Q$-ball can be obtained, by integrating (\ref{dphi}), as the simple function
\be
\phi(r)=\cot^{-1}\left(e^{r^2/2}\cot\phi_0\right)=\arctan\left(e^{(R^2-2r^2)/4}\sqrt{e^{R^2/2}-2}\right)\ ,
\ee
where the last expression is written in terms of the $Q$-ball radius, defined in this case by $\phi(R)=\phi_0/2$.
Figure~\ref{fig:VtV} shows the potential and tunneling potential for the two choices $\phi_0=\pi/4,\pi/2-10^{-7}$, (corresponding to $R=1.33$ and $R=5.68$ respectively), the last one being in the thin-wall regime. The corresponding field profiles are given in the bottom plot.

\subsection{Cubic \texorpdfstring{$\mathcal{E}$}{E}}

Another example is
\be
\mathcal{E}(\phi)=\phi^2(\phi_a-\phi)\,,
\ee
which is the solution for the potential
\be
V(\phi)=-\frac{\phi^2}{4\phi_a}\left[4\phi_a(\phi-\phi_a)+(2\phi_a-3\phi)^2\log
\frac{(2\phi_a-3\phi)\phi_0}{(2\phi_a-3\phi_0)\phi}
\right] .
\ee
The profile of the $Q$-ball can also be obtained as 
\be
\phi(r)=\frac{\phi_a}{3}\left[1-\tanh\left(\phi_a r^2/4-B\right)\right] ,
\ee
with 
\be
B=-\frac12\log\left(\frac{2\phi_a}{3\phi_0}-1\right) ,
\ee 
so that $\phi(0)=\phi_0$. In terms of the $Q$-ball radius $R$, defined by $\phi(R)=\phi_0/2$, the profile can be written as
\be
\phi(r)=\frac{2\phi_a\left(e^{\phi_a R^2/2}-2\right)}{3\left(e^{\phi_a R^2/2}+e^{\phi_a r^2/2}-2\right)}\ .
\ee
This case also admits a thin-wall limit, for $B\gg 1$.

\subsection{Quadratic \texorpdfstring{$\mathcal{E}$}{E}}

The previous two examples (of type A potentials) are simple enough but one does not have the freedom of changing $\omega$ (and therefore $Q$). This can be done in the next example, which is even simpler:
\be
\mathcal{E} = \frac12 m_t^2\phi^2 \ ,\quad
V= \frac12 m_t^2\phi^2 \log(\phi/\phi_c)+\frac12 \omega^2\phi^2\ , 
\label{eq:quadratic_E}
\ee
where $m_t$, $\phi_c$ and $\omega$ are free parameters (actually
$\phi_c$ can be absorbed into $\omega$, but we want $\omega$ to be a separate adjustable parameter). Now we can keep $U(\phi)= -\frac12 m_t^2\phi^2 \log(\phi/\phi_c)$ fixed and vary $\omega$ and this will change the value of
the starting value $\phi_0$, which is given by
\be
\phi_0(\omega) =\phi_c\, e^{1-\omega^2/m_t^2}\ .
\ee
The $Q$-ball profile can also be obtained as
\be
\phi(r)=\phi_0(\omega)\, e^{-m_t^2r^2/4}\ .
\ee
This result reproduces an exact solution that was already discussed in Refs.~\cite{Enqvist:1998en,Copeland:2009as}.
From this profile it is clear that this example does not have a thin-wall limit. Indeed, the value of the potential at
the minimum, $V_M$, is related to the value of the potential at the initial point, $V_0$, as 
\be
V_M=-\frac{1}{4e^3}m_t^2\phi_0^2(\omega)=-\frac{1}{2e^3}V_0\ ,\label{exampleQE}
\ee
and one cannot have $|V_M|\gg \Delta V=V_+-V_0$.
Alternatively, one can notice that $V(\phi)$ does not have a maximum for $\phi >0$.

\begin{figure}[t!]
\begin{center}
\includegraphics[width=0.47\textwidth]{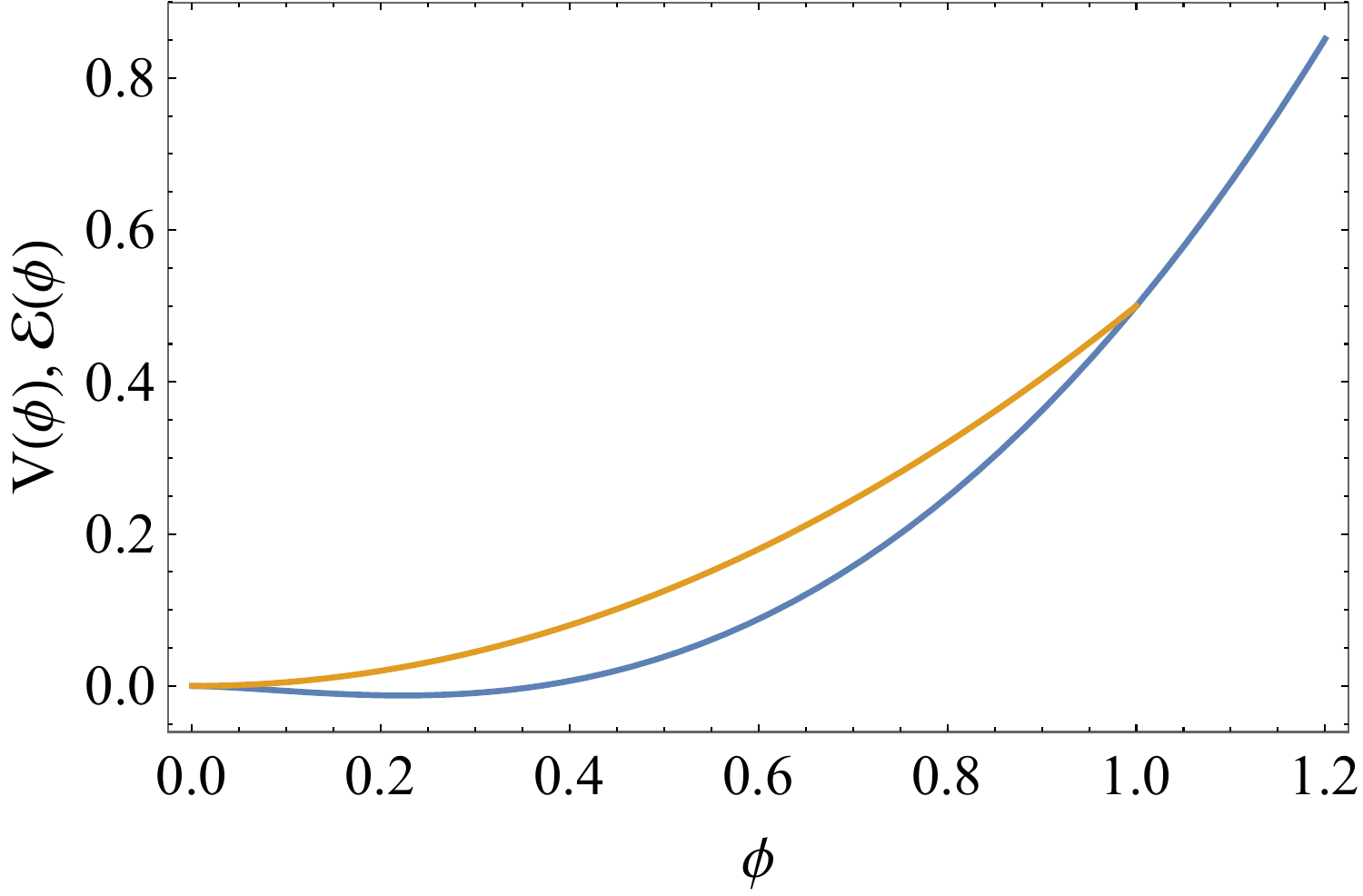}\,
\includegraphics[width=0.46\textwidth]{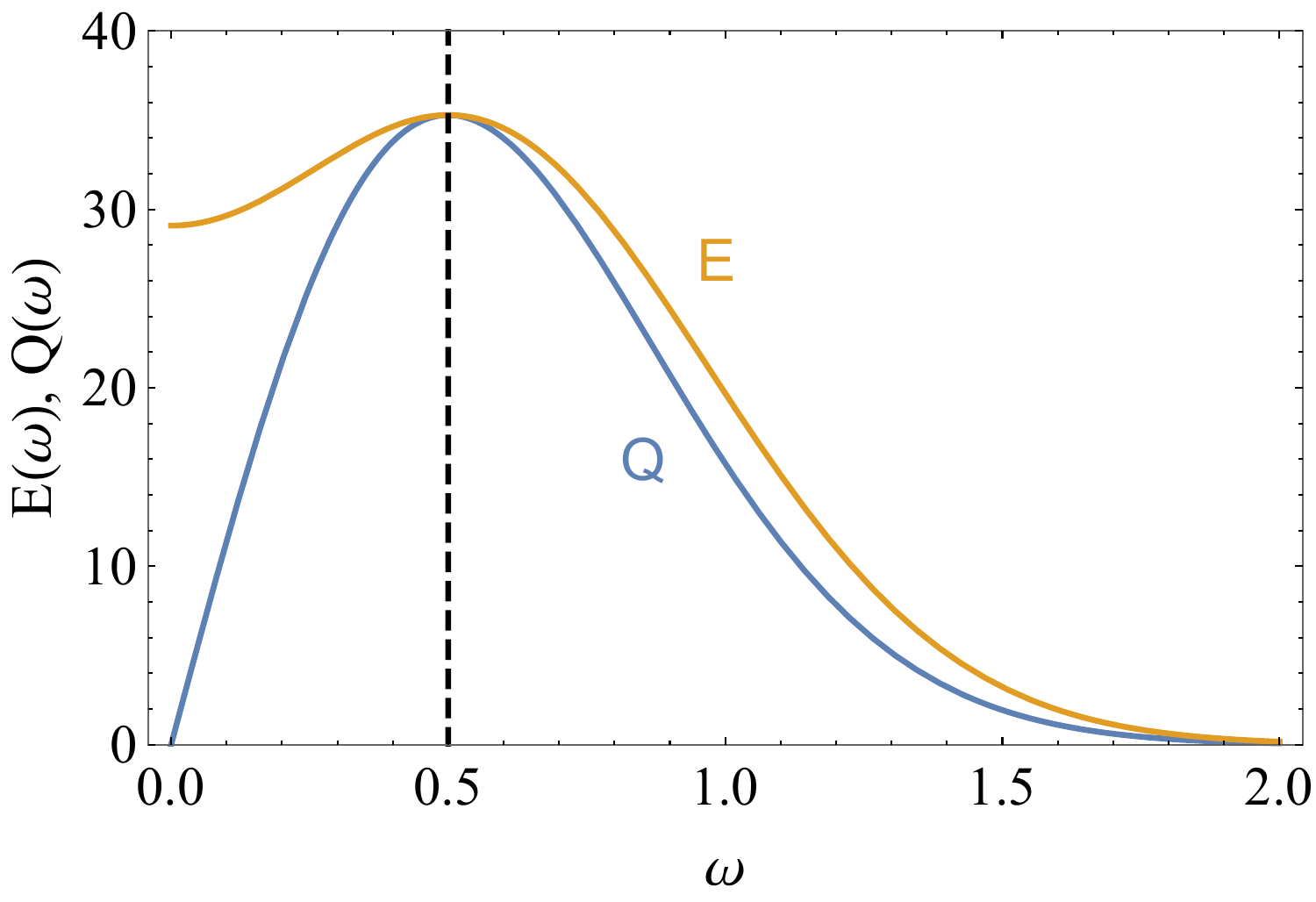}
\end{center}
\caption{For the analytic example of (\ref{eq:quadratic_E}):  Potential, $V$, and tunneling potential, $\mathcal{E}$ for $m_t=\phi_c=\omega=1$,  (left plot) and $Q$-ball charge and energy as functions of $\omega$  for $m_t=\phi_c=1$ (right plot).
\label{fig:QE}
}
\end{figure}

The charge and energy of the $Q$-balls can be calculated analytically as
\be
Q(\omega)=\frac{2\sqrt{2}\pi^{3/2}\omega\phi_0^2(\omega)}{m_t^3}\ , \quad
E(\omega)=\omega Q(\omega)+ \frac{\pi^{3/2}\phi_0^2(\omega)}{\sqrt{2}m_t}\ ,
\ee
which satisfy the identity $dE/d\omega=\omega dQ/d\omega$. Figure~\ref{fig:QE} shows $V$ and $\mathcal{E}$ in this example as well as $Q(\omega)$ and $E(\omega)$ (for $m_t=\phi_c=1$). The dashed line at $\omega=m_t/2$, where $Q$ is maximal, marks the value below which $dQ/d\omega>0$ and the $Q$-balls are unstable.

\subsection{Quartic \texorpdfstring{$\mathcal{E}$}{E}}

In the analysis of the tunneling problem (with $d=4$) there is a simple (scale-independent) potential, $V=-\lambda\phi^4/4$, which admits an infinite family of bounces \cite{Fubini:1976jm}, corresponding to $V_t=-\lambda\phi_0\phi^3/4$ with arbitrary $\phi_0$. 
The tunneling action is $\phi_0$ independent and given by $S=8\pi^2/(3\lambda)$. We can then ask if there is a similar example for $Q$-balls. The formal solution does exist and is given by
\be
V(\phi)=\frac{\lambda\phi^6}{\Lambda^2}\ ,\quad\quad \mathcal{E}(\phi)=\frac{\lambda\phi^4\phi_0^2}{\Lambda^2} \ .
\label{special}
\ee
However, it can be checked that this example is unphysical, as it would give a divergent charge $Q$. Indeed, at small $\phi$, the integrand in (\ref{Qintegral}) goes like $1/\phi^2$, leading to a divergent integral for $\phi\to 0$. 

This divergence can be cured by introducing an additional mass term in $\mathcal{E}$ that acts as an infrared cutoff (at $\phi\to 0$). We take then
\be
\mathcal{E}(\phi) = 2\omega^2\phi^2+\frac{\lambda\phi^4\phi_0^2}{\Lambda^2} \ ,
\ee
where we have written the mass in terms of $\omega$ for later convenience.
For this $\mathcal{E}$ we get
\be
V(\phi)=2\omega^2\phi^2+\frac{\lambda\phi_0^2\phi^4}{\Lambda^2}+\omega^2\left(1+C \phi^2/\phi_0^2\right)^2\phi^2\log\frac{(1+C)\phi^2}{\phi_0^2+C\phi^2}\ ,
\ee
where $C\equiv \lambda\phi_0^4/(\Lambda^2\omega^2)$.
The $Q$-ball profile is 
\be
\phi(r)=\frac{\phi_0}{\sqrt{(C+1)e^{2\omega^2r^2}-C}}\ .
\ee
For $\omega/\phi_0\ll 1$, this potential is well approximated, at large $\phi$, as
\be
V(\phi)\simeq \left(\frac{\lambda}{\Lambda^2}-\frac{\omega^2}{2\phi_0^4}\right)\phi^6+\frac{2\omega^2}{\phi_0^2}\phi^4+\frac12\omega^2\phi^2
\simeq \frac{\lambda\phi^6}{\Lambda^2}+\frac12\omega^2\phi^2\ ,
\ee
which is independent of $\phi_0$, has an adjustable $\omega$ parameter and 
takes the form (\ref{special}) up to a small IR cutoff correction.
For small $\phi$, one has
\be
V(\phi)\simeq \omega^2\left[2+\log(1+C)+2\log(\phi/\phi_0)\right]\phi^2\ .
\ee
This form leads to the existence of a shallow minimum at
\be
\phi_m\simeq \frac{\phi_0}{e^{3/2}\sqrt{1+C}}\simeq \frac{\omega\Lambda}{e^{3/2}\sqrt{\lambda}\phi_0}\ .
\ee
Nevertheless, the parameter $\omega$ only approximately works as it should. In particular, the relation $dE/d\omega=\omega dQ/d\omega$
breaks down as $\omega\to 0$, although it holds for larger $\omega$ when $C\ll1$.

\section{Conclusion \label{sec:Conclusion}} 

In this article we used the mathematical similarity between $Q$-balls and vacuum-decay bounces to reformulate the $Q$-ball problem in the \emph{tunneling potential approach} of Ref.~\cite{Espinosa:2018hue}. Rather than tracking $\phi(r)$ -- or particle position $\phi$ as a function of time $r$ in the rolling-particle analogy -- we can track the particle's energy $\mathcal{E}$ as a function of position, $\mathcal{E}(\phi)$. Since the latter is monotonically decreasing due to the friction term in the equation of motion these two are equivalent formulations.
One advantage of the $\mathcal{E}(\phi)$ formulation is that $\mathcal{E}(\phi)$ can be approximated fairly accurately with low-order polynomials, providing simple numerical estimates for $Q$-ball properties that work especially well in the small $Q$-ball regime that is usually inaccessible via approximations. We have shown that this works very well for essentially all kinds of single-field $Q$-ball potentials.
Furthermore, the $\mathcal{E}(\phi)$ formulation provides a new angle to find exactly solvable $Q$-ball potentials, as we have illustrated with several examples.

\section*{Acknowledgements}

The work of JH and MS was supported in part by the National Science Foundation under Grant No.~PHY-2210428. The work of JRE has been funded by the following grants: IFT Centro de Excelencia Severo Ochoa SEV-2016-0597, CEX2020-001007-S and by PID2019-110058GB-C22 funded by MCIN/AEI/10.13039/501100011033 and by ``ERDF A way of making Europe''.

\appendix

\section{Large \texorpdfstring{$\bma{Q}$}{Q}-Balls in Flat Potentials \label{app:flat}} 

In this appendix we derive the large $Q$-ball behavior for the flat potential case, as approximated through the exactly solvable potential of Eq.~\eqref{eq:piecewise_quadratic}, repeated here for the convenience of the reader:
\begin{align*}
U(\phi) = \begin{cases}
m_\phi^2 |\phi|^2 \,, & \text{ for } |\phi(r)|\leq \Lambda\,,\\
m_\phi^2 \Lambda^2 \,, & \text{ for } |\phi(r)|> \Lambda\,.
\end{cases}
\end{align*}
 By rescaling the field $( \phi(\Vec{x},t) = \Lambda e^{i \omega t} f(|\Vec{x}|) )$ and spatial coordinate $( \rho \equiv m_{\phi} |\Vec{x}| )$ we arrive at the differential equation
\begin{align}
f''(\rho) +\frac{2}{\rho} f'(\rho)  = 
\begin{cases}
(1-\kappa^2) f(\rho) \,, & f \leq 1\,,\\
-\kappa^2 f(\rho) \,, & f >1\,,
\end{cases}
\label{eq:piecewise_eom}
\end{align}
where $\kappa \equiv \omega/m_{\phi}$ is the only remaining variable, with values $\kappa \in (0,1)$. Demanding the solution to be continuous and differentiable, we find
\begin{align}
    f = \begin{cases}
        f(0)\,\frac{\sin (\kappa  \rho )}{\kappa  \rho }\,, & \rho < R\,,\\
        f(0)\,\frac{\sin (\kappa R)}{\kappa}\frac{1}{\rho }\, \exp\left(\sqrt{1-\kappa ^2} (R-\rho )\right) , &\rho\geq R\,,
    \end{cases}
\end{align}
where the matching radius is given by
\begin{align}
    R= \frac{\pi -\arcsin{\kappa}}{\kappa} = \frac{\pi}{\kappa} - 1 -\mathcal{O}(\kappa^2)\,.
\end{align}
The overall prefactor $f(0)$ is not fixed by the linear differential equation but its $\omega$ or $\kappa$ dependence can be obtained by demanding $dE/d\omega = \omega dQ/d\omega$ to hold, $E$ and $Q$ being calculated through the integrals
\begin{align}
\begin{split}
    &\int\limits_0^\infty \dd \rho\, \rho^2 f^2 =f(0)^2\frac{ \left(2 \cos ^{-1}(\kappa )+\pi +\frac{2 \kappa }{\sqrt{1-\kappa ^2}}\right)}{4 \kappa ^3}\,,\\
    &\int\limits_0^\infty \dd \rho\, \rho^2 f'^2 =f(0)^2\frac{ \left(2 \cos ^{-1}(\kappa )+\pi \right)}{4 \kappa }\,.
\end{split}
\label{eq:flat_integrals}
\end{align}
$dE/d\omega = \omega dQ/d\omega$ can be translated into the relation
\begin{align}
    \int\dd\rho \rho^2 f^2 = -\frac{1}{3\kappa}\frac{\dd}{\dd\kappa} \int\dd\rho \rho^2 (f')^2\,,
\end{align}
which is satisfied by $f(0) = c\, (\pi -\arcsin{\kappa})/\kappa $, where $c$ is a $\kappa$-independent constant that is fixed to $c=1$ by
 calculating the energy lost to friction and equating it to the potential difference~\cite{Mai:2012cx,Heeck:2020bau},
\begin{align}
    V(0)-V(f(0))=-\int \dd\rho \, \frac{2}{\rho}  (f')^2\,.
\end{align}
We hence find the relations $f(0)= R$ and also $f(R) = 1$, which complete the solution to our differential equation~\eqref{eq:piecewise_eom}. Energy and charge are then given by the integrals in Eq.~\eqref{eq:flat_integrals}.
In particular, we find that these solitons are stable, $E < m_\phi Q$,  for $\kappa < 0.84$, i.e.~once their charge exceeds the critical value
\begin{align}
    Q_{\text{crit}} \simeq 435.44\, \frac{\Lambda ^2}{ m_\phi ^2 } \,.
\end{align}
Our classical-field-theory calculation of course breaks down for small $Q$ and needs to be replaced by a quantum-mechanical treatment, beyond the scope of this article. 
Notice that the physical radius satisfies $R \gtrsim 1/m_\phi$, in agreement with Ref.~\cite{Freivogel:2019mtr}.

Let us focus on the large $Q$-ball case, since this is the region where the above solution is a good approximation to the continuous potential of interest in the main text. For small $\omega$, we have
\begin{align}
Q &\simeq \frac{4\pi^4 m_\phi^2 \Lambda^2}{\omega^4}\left[ 1	- \frac{2\omega}{m_\phi \pi} + \frac{\omega^2}{m_\phi^2 \pi^2} + \mathcal{O}\left(\frac{\omega^4}{m_\phi^4}\right)\right]
\end{align}
and
\begin{align}
E(Q) &\simeq \frac{4 \pi  }{3} \sqrt{2 m_\phi\Lambda}\, Q^{3/4}	- 2\pi \Lambda Q^{1/2} + \mathcal{O}\left(Q^{1/4}\right).
\end{align}
While the general scaling with $\omega$ agrees with the literature~\cite{Dvali:1997qv,MacKenzie:2001av} we provide prefactors and higher-order corrections here.

\bibliographystyle{utcaps_mod}
\bibliography{BIB}

\end{document}